\documentclass[reprint,aps,pre]{revtex4-1}
\usepackage{graphicx}
\usepackage{dcolumn}
\usepackage{amsmath}
\usepackage{amssymb}
\usepackage{amsfonts}
\usepackage{bm}
\usepackage{color}
\usepackage{tikz}
\usepackage{atbegshi}
\usepackage{subfigure}
\usepackage{wasysym}
\usetikzlibrary{shapes,patterns,arrows}
\newcommand{\be}{\begin{equation}}
\newcommand{\ee}{\end{equation}}
\newcommand{\ba}{\begin{eqnarray}}
\newcommand{\ea}{\end{eqnarray}}

\begin{document}

\author{Gianmarco Muna\`o$^1$}
\email{gmunao@unisa.it}
\author{Andrea Correa$^2$}
\author{Antonio Pizzirusso$^1$}
\author{Giuseppe Milano$^{1,3}$}
\affiliation{
$^{1}$Dipartimento di Chimica e Biologia, Universit\`a degli Studi di Salerno,
Via Giovanni Paolo II, 84084 Salerno, Italy. \\
$^{2}$Department of Chemical Science, Federico II University of Naples, via Cinthia,
Complesso Monte S.Angelo, 80126 Napoli, Italy. \\
$^{3}$Department of Organic Materials Science, University of Yamagata, 
4-3-16 Jonan Yonezawa, Yamagata-ken 992-8510, Japan. \\ 
}

\title{On the calculation of potential of mean force between atomistic 
nanoparticles}

\begin{abstract}
We study the potential of mean force (PMF) between atomistic silica and gold
nanoparticles in the vacuum by using molecular dynamics simulations. 
Such an investigation is devised in order to fully characterize the effective interactions
between atomistic nanoparticles, a crucial step to describe the PMF in high-density coarse-grained
polymer nanocomposites.
In our study, we first investigate the behavior of silica nanoparticles, 
considering cases corresponding to different particle sizes and assessing results against
an analytic theory developed by Hamaker for a system of Lennard-Jones
interacting particles [H. C. Hamaker, {\it Physica A}, 1937, {\bf 4}, 1058]. 
Once validated the procedure, we calculate effective interactions between 
gold nanoparticles, which are considered both bare and coated with polyethylene chains,
in order to investigate the effects of the grafting density $\rho_g$ 
on the PMF. 
Upon performing atomistic molecular dynamics simulations, it turns out
that silica nanoparticles experience similar interactions regardless of the particle size, 
the most remarkable difference being a peak in the PMF due to 
surface interactions, clearly apparent for the larger size. 
As for bare gold nanoparticles, they are slightly interacting, the strength of the
effective force increasing for the coated cases. The profile of the
resulting PMF resembles a Lennard-Jones potentials for intermediate
$\rho_g$, becoming progressively more repulsive for high $\rho_g$ and low interparticle
separations. 

\end{abstract}

\maketitle

\section{Introduction}
The potential of mean force (PMF) is one of the most topical issues 
when facing the problem of determining the stability
of nanoparticle (NP) systems and nanocomposites~\cite{Akcora:09,Kawada:17,Baran:17}. The possibility to
calculate the net interactions between a couple of nanoparticles, hence gaining knowledge
on the overall behavior of the system, can open the way to systematic studies of macroscopic
properties of potential tecnological interest. This is particularly true for polymer 
nanocomposites, where the addition of nanoparticles can sensibly improve their physico-chemical
properties (see, {\it e.g.}, Refs.~\cite{Akcora:10,Han:11,Kim:12}). For instance,
it is now well established that a specific NP dispersion state in a polymer matrix is
crucial to improve a given property of the system~\cite{Leibler:02,Kumar:13}; the knowledge of 
such a dispersion state
can be gained by means of the PMF between the NPs belonging to the composite. In this context
it is worth noting that the study of interactions between nanoparticles dispersed 
in polymer matrices or in solvents 
is currently object of rather extensive studies, by means of experimental~\cite{Chevigny:11,You:17},
theoretical~\cite{Martin:13,Ganesan:14} and simulations~\cite{Meng:12,Karatrantos:17} 
approaches. 

As far as theoretical approaches are concerned, they are generally 
based on the Polymer Reference Interaction Site Model (PRISM) theory
developed by Curro and Schweizer in the late '80s~\cite{Curro1,Curro2}. 
This theory has 
generally provided good results when facing the study of PMF, obaining a good agreement with
experimental data
(see, {\it e.g.} Refs.~\cite{Hooper:04,Jayaraman:09} and~\cite{Raos:08,Ganesan:10} for two
detailed reviews); however,
according to this theory, only generic models can be investigated and the
chemical details characterizing a given compound is lost. In order to fully
recover its chemical structure, computer-simulations based
calculations are needed. Many efforts have been dedicated to shed light on
the complex properties of polymers-nanoparticles interface, going from
atomistic to mesoscale representations (see, for instance, 
Refs.~\cite{Karatrantos:16,Kumar:17} for two recent reviews). 
To quote some examples, previous numerical studies have highlighted the
effects of the NP curvature on the behavior of PMF~\cite{Cerda:03},
the role played by attractive dispersions interactions between polymer and
nanoparticles~\cite{Smith:03,Marla:06} and the importance of the ratio between
coated and free polymer chain lengths~\cite{Smith:09} and of the NP
radius~\cite{Loverso:11}.

\begin{figure*}
\begin{center}
\begin{tabular}{ccc}
\includegraphics[width=5.5cm,angle=0]{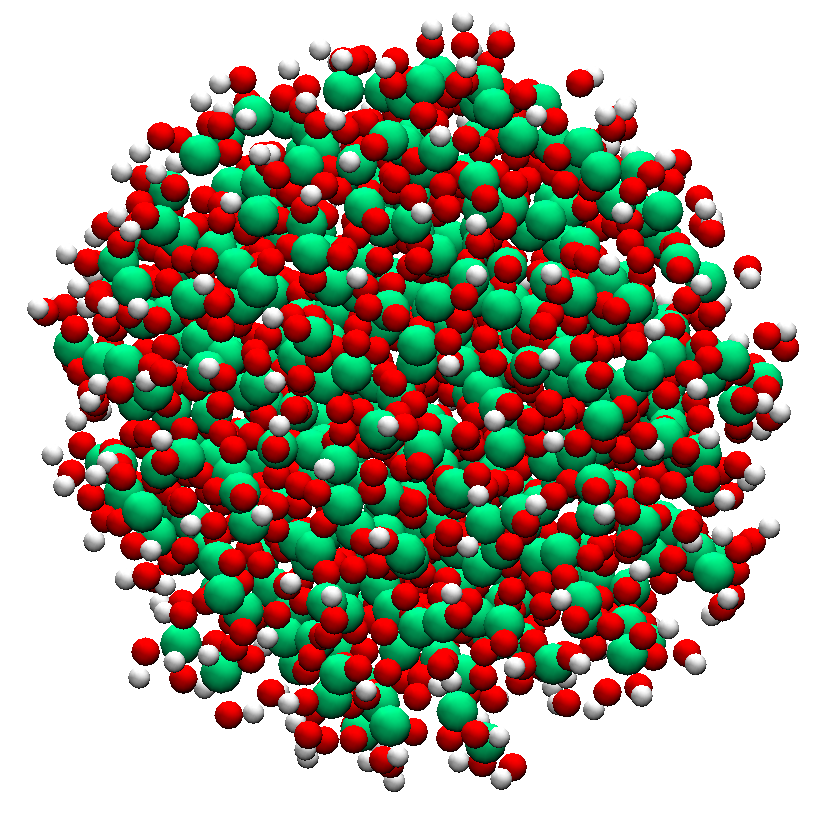} \qquad
\includegraphics[width=3.0cm,angle=0]{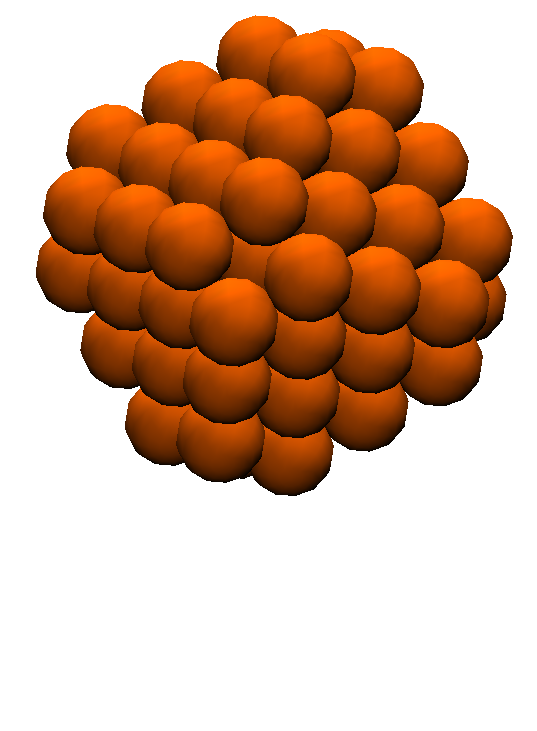} \qquad
\includegraphics[width=7.0cm,angle=0]{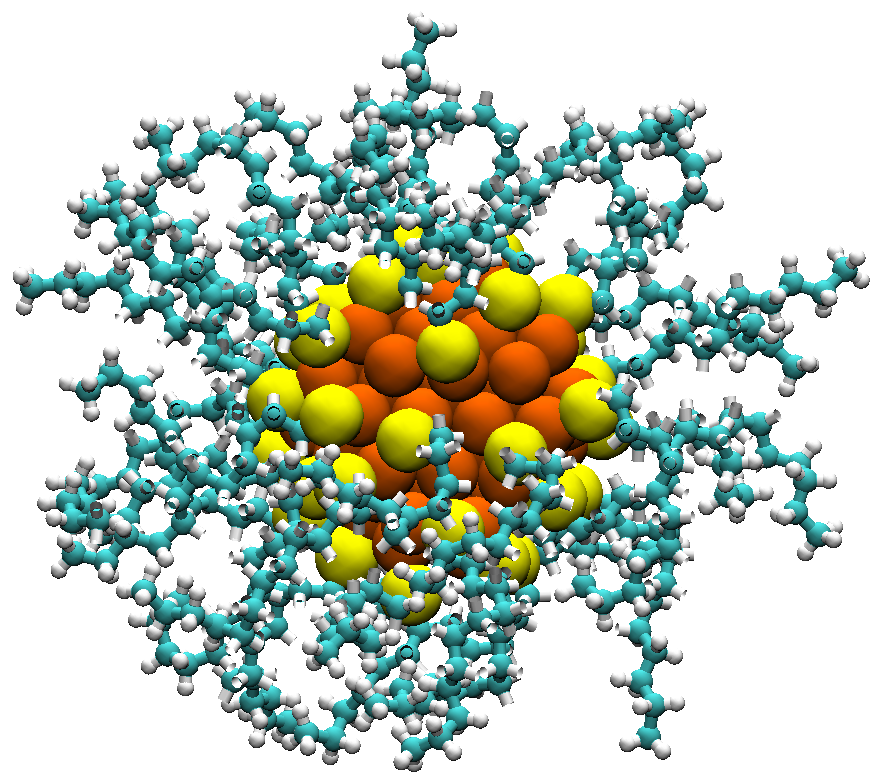} 
\end{tabular}
\caption{Representative snapshots of nanoparticles investigated in this 
work: silica NP (left), bare gold NP (center) and coated gold NP (right).
The color legend is the following: O=red, Si=green, H=white, Au=orange, S=yellow, C=cyan.}
\label{fig:snap}
\end{center}
\end{figure*}
In all the above mentioned studies, nanoparticles have been generally represented
by means of simple coarse grained models, like bead-spring or pearl-necklace representations; 
this is an unavoidable choice when tackling
the issue of investigating nanoparticles dispersed in polymer matrices. In fact,
under such conditions, an atomistic representation should require the calculation
of pairwise interactions over hundreds of thousands of atoms for long times,
making the calculation practically unaffordable. More details on the chemical structure
of the coarse-grained models can be gained by adopting a mean field representation
of the non-bonded interactions, hence giving rise to an hybrid particle-field
approach~\cite{Milano:09,Milano:10}. However, the atomistic details of the interactions
equally need to be addressed in order to provide the overall behavior of the potential 
of mean force.

In this work we plan to address this point by calculating the PMF between
couples of silica and gold nanoparticles by using atomistic models. 
In particular, we focus on the nature and a suitable procedure to obtain 
realistic NP-NP interaction potentials.
Our investigation is focused on the effects due to the silica particle size
and to the presence (or absence)
of polyethylene (PE) chains coated onto the surface of gold nanoparticles. 
Such models have been chosen in order to
consider systems with an increasing degree of complexity, from bare
nanoparticles to coated systems with increasing grafting densities.
In order to
make the calculation affordable, and aiming to clarify the microscopic 
details of NP-NP interactions,
we simulate our systems in the vacuum, {\it i.e.} without any surrounding polymer matrix or solvent.
The present work is framed in the broader perspective of obtaining atomistic potentials that 
will be used in subsequent studies concerning high-density coarse-grained polymer nanocomposites.
For such an aim, we perform molecular dynamics
simulations in the canonical ensemble (NVT) by means of the GROMACS
package~\cite{Gromacs:08}. 
As stated in Ref.~\cite{Meng:12}, when compared to 
other techniques frequently adopted for calculating the PMF, like the
umbrella sampling, this approach shows some advantages: in fact, even if the results obtained
by implementing the two techniques are basically the same, NVT simulations
allow to
gain information also on the net forces experienced by the nanoparticles, 
and to discriminate between their different contributions.
For the case of silica nanoparticles we also compare simulation
results with theoretical predictions based on the Hamaker theory~\cite{Hamaker:37},
specifically developed to deal with spherical systems comprised by several Lennard-Jones interacting
particles. Such a comparison is not performed for the gold NPs, due to the presence of
coating chains and the non-spherical shape of the particles. 
In addition, in order to gain further insigth into the local structure
of coated PE chains, gyration radii and end-to-end distances have been also computed. 

The paper is organized as follow: in the next Section we provide details on models, theory
and simulation approaches. In
Section III we present and discuss the obtained results, finally drawing the
conclusions in the last Section.
 
\section{Models and methods}
\subsection{Silica and gold NPs}
\begin{table}
\begin{center}
\caption{NP systems investigated in this work. Diameters are in nm and grafting densities $\rho_g$ in
chains/nm$^2$. In the case of gold nanoparticles, the diameter is referred to the circumscribed 
sphere of the cuboctahedron.}\label{tab:NP}
\begin{tabular*}{0.45\textwidth}{@{\extracolsep{\fill}}cccccccccccccc}
\hline
\hline
& NP & \qquad & Diameter & \qquad & Coated  & \qquad & $N_a$ 
& \qquad & $\rho_g$ \qquad & $N_c$ \qquad & $L_c$ \\
\hline
& Silica & \qquad & 2.5 & \qquad & No & \qquad & 1401 & \qquad & -  \qquad & -  \qquad & -\\
& Silica & \qquad & 4.0 & \qquad & No & \qquad & 3189 & \qquad & -  \qquad & -  \qquad & -\\
& Gold & \qquad & 1.6 & \qquad & No & \qquad & 79 & \qquad & - \qquad & - \qquad & - \\
& Gold & \qquad & 1.6 & \qquad & Yes & \qquad & 801 & \qquad & 2.36 \qquad & 19 \qquad & 38\\
& Gold & \qquad & 1.6 & \qquad & Yes & \qquad & 1143 & \qquad & 3.48 \qquad & 28 \qquad & 38 \\
& Gold & \qquad & 1.6 & \qquad & Yes & \qquad & 1523 & \qquad & 4.72 \qquad & 38 \qquad & 38 \\
\hline
\end{tabular*}
\end{center}
\end{table}
\begin{table}
\begin{center}
\caption{Parameters of non-bonded potential 
$V_{nb}(r_{ij})\equiv V_{LJ}(r_{ij})+V_{Coul}(r_{ij})+V_{rf}(r_{ij})$.}\label{tab:pot}
\begin{tabular*}{0.45\textwidth}{@{\extracolsep{\fill}}cccccccc}
\hline
\hline
& Atom & \qquad & $\sigma$ (nm) & \qquad & $\epsilon$(kJ mol$^{-1})$  & \qquad
& $q$(e) \\
\hline
& Silica NP & \qquad & & \qquad & \\
& Si & \qquad & 0.392000 & \qquad & 2.510400 & \qquad & 1.020\\
& O & \qquad & 0.315400 & \qquad & 0.636840 & \qquad & -0.510\\
& H & \qquad & 0.235200 & \qquad & 0.092000 & \qquad & 0.255\\
& Gold NP & \qquad & & \qquad & \\
& Au & \qquad & 0.293373 & \qquad & 0.163176 & \qquad & 0.000\\
& S & \qquad & 0.355000 & \qquad & 1.066000 & \qquad & -0.180\\
& C & \qquad & 0.350000 & \qquad & 0.276000 & \qquad & -0.120\\
& H & \qquad & 0.250000 & \qquad & 0.138000 & \qquad & 0.060\\
\hline
\end{tabular*}
\end{center}
\end{table}
A representative picture of the nanoparticle models investigated in this 
work is given in Fig.~\ref{fig:snap}: as visible, different colors
label different atom types. 
As anticipated in the Introduction, we first 
investigate the behavior of the PMF between a couple of 
bare spherical silica nanoparticles (Fig.~\ref{fig:snap}, left panel) with 
diameters of 2.5 and 4 nm; the model for these NPs has been developed  
by the M{\"u}eller-Plathe group and employed for studying the interface between 
silica nanoparticles and polymer matrices~\cite{Ndoro:11,Eslami:13}.
Then, we study the behavior of gold NPs (middle panel
of Fig.~\ref{fig:snap}): for such an aim we adopt a model already investigated
in Refs.~\cite{Rai:04,Milano-Gold}, constituted 
by a core made of 79 Au atoms organized in a cuboctahedral
geometry with a Au-Au bond length of 0.292 nm. Finally, gold NPs are 
considered coated with a variable number of PE chains, each one containing 
38 monomers (right panel of Fig.~\ref{fig:snap})
and connected to the inner part of the NP through one sulfur atom.
Following the prescription of Ref.~\cite{Rai:04}, 
sulfur atoms are covalently bonded
to a gold atom by using a harmonic potential with an Au-S bond
length of 0.24 nm (see Tab~\ref{tab:bond}):   
this method of attachment restricts the position of the S atom
to a position directly above the Au atom to which it is attached.
Further details on the model parameters can be found in Refs.~\cite{Rai:04,Milano-Gold}.

\begin{table}
\begin{center}
\caption{Parameters of bond stretching potential 
$V_{b}(r)\equiv (k_r/2)(r-r_0)^2$.}
\label{tab:bond}
\begin{tabular*}{0.45\textwidth}{@{\extracolsep{\fill}}cccccc}
\hline
\hline
& Bond & \qquad & $r_0$ (nm) & \qquad & k$_r$(kJ mol$^{-1}$ nm$^{-2})$ \\
\hline
& Silica NP & \qquad & & \qquad & \\
& Si-O & \qquad & 0.1630 & \qquad & $1 \cdot 10^7$ \\
& O-H & \qquad & 0.0950 & \qquad & $1 \cdot 10^7$ \\
& Gold NP & \qquad & & \qquad & \\
& Au-Au & \qquad & 0.2920 & \qquad & 400000 \\
& Au-S & \qquad & 0.2400 & \qquad & 400000 \\
& C-S & \qquad & 0.1810 & \qquad & 400000 \\
& C-H & \qquad & 0.1090 & \qquad & 400000 \\
& C-C & \qquad & 0.1552 & \qquad & 265265 \\
\hline
\end{tabular*}
\end{center}
\end{table}
The complete collection of the investigated systems
is reported in Tab.~\ref{tab:NP}, where $N_a$ is the total number of atoms,
$\rho_g$ the grafting density, $N_c$ the number of coated chains and $L_c$ the number 
of monomers belonging to a single chain. 
Tables~\ref{tab:pot}-\ref{tab:dih} 
provide a summary of the potential energy parameters.  
In particular in Tab.~\ref{tab:pot} we report all the non-bonded interactions,
i.e. the interactions between atom pairs whose distance $r_{ij}$
is not fixed by the
connectivity. In this Table we have defined:
\begin{equation}
V_{LJ}(r_{ij}) = 4\epsilon[(\sigma/r_{ij})^{12}-(\sigma/r_{ij})^{6}] \,,
\end{equation} 
\begin{equation}
V_{Coul}(r_{ij}) = q_iq_j/4\pi\epsilon_0r_{ij}^2 \,,
\end{equation} 
\begin{equation}
V_{rf}(r_{ij}) = V_{Coul}(r_{ij}) r_{ij} (\epsilon_{rf}-1)
(2\epsilon_{rf}+1) (r_{ij}^2/r_{cut}^3) \,.
\end{equation} 
The first contribution is the standard Lennard-Jones potential, determined
by the interaction energy $\epsilon$ and the close-contact distance $\sigma$.
$V_{Coul}(r_{ij})$ is the Coulombic interaction between two atoms with 
charges $q_i$ and $q_j$, $\epsilon_0$ being the vacuum permittivity. 
The third contribution takes into account the effect of a reaction
field~\cite{Tironi:95} with a dielectric constant $\epsilon_{rf}$ and, after the cutoff length
$r_{cut}$, is modeled by using the Kirkwood approximation~\cite{Kirkwood:35}.
In Tab~\ref{tab:bond} we define the bond stretching potential $V_{b}(r)$,
dependent on the elastic constant $k_r$ of the material and on the
elongation $r$ respect to the equilibrium position $r_0$. Analogous
expressions hold for the bond angle potential $V_a(\theta)$ reported in 
Tab~\ref{tab:ang}, where $\theta$ and $\theta_0$ are the angular
counterparts of $r$ and $r_0$ and $k_{\theta}$ is a constant 
still dependent on the material properties.
Finally, parameters of the dihedral angle potentials
$V_d(\theta)$ for the polyethylene chains are
reported in Tab~\ref{tab:dih}, $\phi$ and $\phi_0$ being the analogous
of $\theta$ and $\theta_0$ for dihedral angles, $k_{\phi}$ an other
material-dependent constant and $f$ being the multiplicity
of $\phi_0$.

\begin{table}
\begin{center}
\caption{Parameters of bond angle potential $V_d(\theta) \equiv (k_{\theta}/2)
(\theta-\theta_0)^2$.}\label{tab:ang}
\begin{tabular*}{0.45\textwidth}{@{\extracolsep{\fill}}cccccc}
\hline
\hline
& Bond angle& \qquad & $\theta_0$(degrees) & \qquad & k$_\theta$(kJ mol$^{-1}$ rad$^{-2})$ \\
\hline
& Silica NP & \qquad & & \qquad & \\
& O-Si-O & \qquad & 109.47 & \qquad & 469.716 \\
& Si-O-Si & \qquad & 144.00 & \qquad & 209.598 \\
& Si-O-H & \qquad & 119.52 & \qquad & 228.836 \\
& Gold NP & \qquad & & \qquad & \\
& H-C-H & \qquad & 107.8 & \qquad & 276.144 \\
& H-C-C & \qquad & 110.7 & \qquad & 292.88 \\
& C-C-C & \qquad & 112.7 & \qquad & 527.184 \\
& C-C-S & \qquad & 108.6 & \qquad & 418.4 \\
& S-C-H & \qquad & 113.4 & \qquad & 292.88 \\
\hline
\end{tabular*}
\end{center}
\end{table}
\begin{table}
\begin{center}
\caption{Parameters of dihedral angle potentials 
$V_a(\phi) \equiv (k_{\phi}/2)
[1-\cos f(\phi - \phi_0)]$.}\label{tab:dih}
\begin{tabular*}{0.45\textwidth}{@{\extracolsep{\fill}}cccccc}
\hline
\hline
& Bond angle& \qquad & $\phi_0$(degrees) & \qquad & k$_\phi$(kJ mol$^{-1}$ rad$^{-2})$ \\
\hline
& Polyethylene & \qquad & & \qquad & \\
& H-C-C-C & \qquad & 60 & \qquad & 5.86 \\
& C-C-C-C & \qquad & 60 & \qquad & 5.86 \\
& C-C-C-S & \qquad & 60 & \qquad & 5.86 \\
\hline
\end{tabular*}
\end{center}
\end{table}

\subsection{Simulation details}
\begin{table}
\begin{center}
\caption{Minimum and maximum values of interparticle distances as functions of particle size
and grafting density. The number of investigated simulation points $N_p$ is also reported.
All distances are in nm. As stated in Tab.~\ref{tab:NP}, for
gold nanoparticles, the diameter is referred to the circumscribed sphere of the 
cuboctahedron.}\label{tab:sim}
\begin{tabular*}{0.45\textwidth}{@{\extracolsep{\fill}}cccccccccccccc}
\hline
\hline
& NP & \qquad & Diameter & \qquad & Coated  & \qquad & $\rho_g$ 
& \qquad & $r_{min}$ \qquad & $r_{max}$ \qquad & $N_p$ \\
\hline
& Silica & \qquad & 2.5 & \qquad & No & \qquad & - & \qquad & 2.5  \qquad & 8.5  \qquad & 31 \\
& Silica & \qquad & 4.0 & \qquad & No & \qquad & - & \qquad & 4.0  \qquad & 10  \qquad & 36\\
& Gold & \qquad & 1.6 & \qquad & No & \qquad & - & \qquad & 1.6 \qquad & 5.6 \qquad & 21 \\
& Gold & \qquad & 1.6 & \qquad & Yes & \qquad & 2.36 & \qquad & 1.6 \qquad & 6.6 \qquad & 26\\
& Gold & \qquad & 1.6 & \qquad & Yes & \qquad & 3.48 & \qquad & 1.6 \qquad & 7.6 \qquad & 31 \\
& Gold & \qquad & 1.6 & \qquad & Yes & \qquad & 4.72 & \qquad & 1.6 \qquad & 7.6 \qquad & 31 \\
\hline
\end{tabular*}
\end{center}
\end{table}
In the present work all simulations have been performed by using GROMACS 
4.6.3.~\cite{Gromacs:08},
employing a cubic simulation box of side $L_{box}=26$ nm with periodic boundary
conditions. 
In the case of silica NPs, a time step of 1 fs has
been used for all simulations. For the nonbonded interactions, a
cutoff of 1.0 nm has been used, while the coulomb long-range
electrostatic interactions have been treated by means of a generalized
reaction field~\cite{Tironi:95} with a dielectric constant $\epsilon_{rf}=6.23$ and a cutoff
of 1.0 nm. 
For gold NPs we have analogously proceeded,
the only differences being the values of cutoff for nonbonded and electrostatic 
interactions, both fixed to 1.35 nm. 
Simulation parameters have been fixed
by following a similar procedure described in Ref.~\cite{DeNicola-JPCB}.
In all systems the temperature has been kept
constant at $T= 590 K$ by using a Berendsen thermostat~\cite{Berendsen:84} with a
time coupling $\tau = 0.1$ ps. 
We have verified that results do not change if the Nose-Hoover thermostat
is used after the equilibration in place of the Berendsen one.  
The temperature has been chosen high enough to 
allow for a proper relaxation of the considered systems and for the subsequent PMF 
calculations. The specific value of 590 $K$ has been fixed in order to
simulate silica nanoparticles in conditions similar to those reported in previous numerical 
investigations of the same NPs~\cite{Ndoro:11,Eslami:13}. With the aim to investigate,
for the sake of clarity, silica and gold nanoparticles at the same temperature, 
we have set $T=590 K$ for studying gold NPs also.
In order to calculate the PMF between the above said nanoparticles, we have
preliminarly built a collection of independent initial configurations with
particles placed at progressively increasing mutual distances. 
In the case of gold coated NPs, we have first prepared configurations with 
the higher grafting density considered in this work (i.e. $\rho_g=4.72$) and
taken from Ref.~\cite{Milano-Gold}: in such configurations, chains are
stretched and uniformly distributed over the NP surface. Configurations with
lower $\rho_g$ have been obtained by deleting some chains, in order to get
a final $N_c$ equal to 50\% ($N_c=19$) or to $\simeq$ 75\% ($N_c=28$) of the
fully coated case.  
All initial systems
have been built by using the Packmol program~\cite{Packmol}, which allows one
to put the desired number of particles in a given position 
inside the simulation box avoiding 
overlaps. Then, we have
computed the forces experienced by the nanoparticles, whose centers of mass
are kept fixed, finally
evaluating the PMF by integrating the obtained forces:
\begin{equation}\label{eq:PMF}
U(r)=-\int_{r_{min}}^{r_{max}} F(r) dr
\end{equation}
where $U(r)$ is the PMF, $F(r)$ is the force and $r_{max}$ and $r_{min}$
are the maxim and minimum distances between the NPs, respectively.
In all simulations $r_{min}$ corresponds to the NP diameter (see Tab.~\ref{tab:NP}), while $r_{max}$
indicates a NP-NP distance where the potential can be confidently assumed equal to zero; 
distances are sampled with a step of 0.2 nm. Values of $r_{min}$ and $r_{max}$, along with
the number $N_p$ of points simulated in a single run, are collectively reported 
in Tab.~\ref{tab:sim}.
After a minimization procedure of 15 ps, equilibration runs of 20 ns have
been preliminarly produced, then averaging the forces over the next 10 ns.
The convergence has been ensured by verifying that the average values of the
forces do not change anymore up to the first significant figure.
Standard deviations have been calculated in the production run
according to the formula:
\begin{equation}
s=\sqrt\frac{\sum_{i=1}^N (F_i-\bar{F})^2}{(N-1)}
\end{equation}
where $F_i$ and $\bar{F}$ are respectively the instantaneous and the average
value of the force experienced by the NPs and $N$ is the number of
simulation time steps. An analogous procedure has been implemented in order to calculate
standard deviations for gyration radii and end-to-end distances.
In what follows, if not explicitly reported in the figures, error bars
corresponding to standard deviations are smaller than symbol sizes 
of the corresponding curves.

\subsection{Hamaker theory}
It is known that for molecules containing a large number of atoms experiencing
pair interactions, the
evaluation of the overall potential has an high computational cost,
since it amounts to calculate a double summation over all the interaction 
sites. For a couple of NPs, such a summation $U_{sum}$ 
is written as~\cite{Everaers:03}:
\begin{equation}
U_{sum}=\sum_{i\in NP_1} \sum_{j\in NP_2} U(r_{ij})
\end{equation}
where $U(r_{ij})$ is the pairwise potential. For particles with simple 
geometrical shapes and number density $\rho_i(r)$ of interaction sites, this 
relation can be generalized to a continuum approximation as:
\begin{equation}\label{eq:int}
U_{sum}=\int_{NP_1} \int_{ NP_2} \rho_1(r) \rho_2(r') U(r-r') dV dV'
\end{equation}
where $V$ is the volume of the NP. For two spheres of radius $r_1 \leq r_2$,
volume $V_i=(4\pi/3)r_i^3$, placed at a distance $r_{12} > r_1 + r_2$ and 
containing particles interacting via a Lennard-Jones potential, 
eq.~\ref{eq:int} can be solved by the Hamaker theory~\cite{Hamaker:37}.  
Within this approach, the attractive part of the interaction can be written as:
\begin{eqnarray}\label{eq:att}
U_A & = & -\frac{A_{12}}{6}\Biggl[\frac{2r_1r_2}{r_{12}^2-(r_1+r_2)^2}
+ \frac{2r_1r_2}{r_{12}^2-(r_1-r_2)^2} 
\nonumber\\[4pt]  
& & \quad +{\rm ln} \left(\frac{r_{12}^2-(r_1+r_2)^2}{r_{12}^2-(r_1-r_2)^2}\right) 
\Biggr] 
\end{eqnarray}
where $A_{12}$ is the Hamaker constant and is takes the value
$A_{12}=4\pi^2\epsilon(\rho\sigma^3)^2$, $\epsilon$ and $\sigma$ being the Lennard-Jones parameters
and $\rho$ being the density. The repulsive part of the interaction
can be written as:
\begin{eqnarray}\label{eq:rep}
U_R & = & \frac{A_{12}}{37800}\frac{\sigma^6}{r_{12}} \Biggl [\frac
{r_{12}^2-7r_{12}(r_1+r_2)+6(a_1^2+7a_1a_2+a_2^2)}{(r_{12}-r_1-r_2)^7}
\nonumber\\[4pt]  
& & \quad +\frac
{r_{12}^2+7r_{12}(r_1+r_2)+6(a_1^2+7a_1a_2+a_2^2)}{(r_{12}+r_1+r_2)^7}
\nonumber\\[4pt]
& & \quad -\frac
{r_{12}^2+7r_{12}(r_1-r_2)+6(a_1^2-7a_1a_2+a_2^2)}{(r_{12}+r_1-r_2)^7}
\nonumber\\[4pt]
& & \quad -\frac
{r_{12}^2-7r_{12}(r_1-r_2)+6(a_1^2-7a_1a_2+a_2^2)}{(r_{12}-r_1+r_2)^7}
\end{eqnarray}
By combining Eqs.~\ref{eq:att}-\ref{eq:rep} one can obtain the total 
interaction. Even if the Hamaker theory strictly holds for spherical particles 
interacting via a Lennard-Jones potential only, it can provide a useful
benchmark against which simulation results can be assessed. 
\section{Results and discussion}
\subsection{Silica nanoparticles}
\begin{figure}
\begin{center}
\includegraphics[width=8.0cm,angle=0]{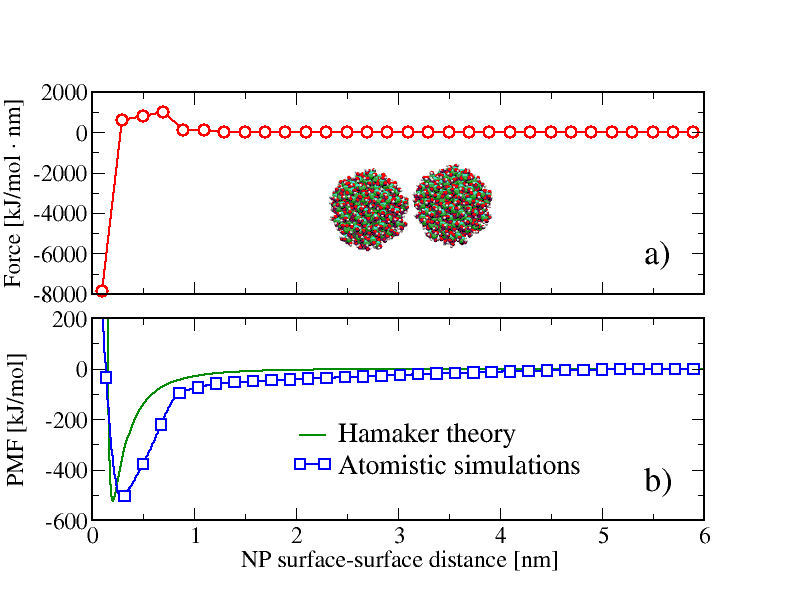} 
\caption{Force (a) and PMF (b) between a pair of silica nanoparticles of diameter 2.5 nm
obtained from atomistic simulations (symbols). In panel (b) a comparison
with the Hamaker theory (full line) is reported. 
}
\label{fig:Hamaker-r1}
\end{center}
\end{figure}

We first calculate forces and PMF between silica nanoparticles:
the absence of coated chains and the contemporary presence of many atoms 
in a single spherical nanoparticle (see Tab.~\ref{tab:NP})
make these NPs ideal candidates in order to make a comparison with the Hamaker theory. 
In order to perform the summation of the bead-bead interactions required by
the theory, we have employed Lennard-Jones parameters for silicon and oxygen atoms taken 
from Tab.~\ref{tab:pot}, then using the Lorentz-Berthelot mixing rules.
Forces and PMF between a pair of silica nanoparticles with radius 2.5 nm are
respectively reported in panels (a) and (b) of Fig.~\ref{fig:Hamaker-r1}, 
along with the predictions due to the Hamaker theory. A pictorial view of two silica NPs
whose mutual distance corresponds to the minimum of the PMF is reported in the snapshot of
panel (a). We first note that the force is strongly negative
for very short interparticle distances; then,
the force shows a steep increase, first attaining positive values and then going to
zero for interparticle distances of $\simeq$ 1.5 nm. 
As for the PMF (panel b), we note that
simulation results closely match the theoretical datum in providing a Lennard-Jones
behavior: in particular the potential shows a well defined minimum of $\simeq$ -500 kJ/mol
observed for a NP surface-surface distance of $\simeq$ 0.3 nm. 
The attractive well is followed by a quick rise,
with the potential going to zero for interparticle separations of $\simeq$ 3 nm. The 
theory slightly anticipates this trend, proving a steeper shape of the potential. 

\begin{figure}
\begin{center}
\includegraphics[width=8.0cm,angle=0]{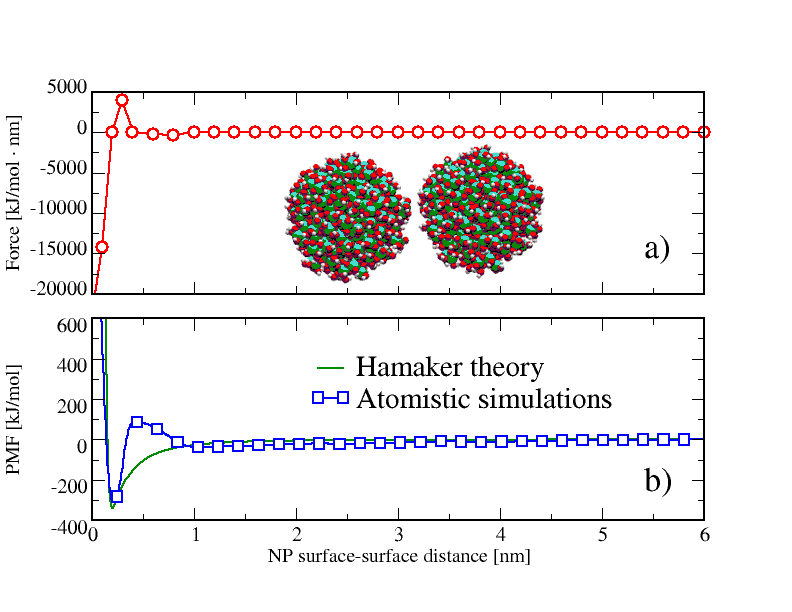} 
\caption{Force (a) and PMF (b) between a pair of silica nanoparticles of diameter 4.0 nm
obtained from atomistic simulations (symbols). In panel (b) a comparison
with the Hamaker theory (full line) is reported. 
}
\label{fig:Hamaker-r2}
\end{center}
\end{figure}
Such a scenario
is not significantly changed upon increasing the size of silica nanoparticles: in 
Fig.~\ref{fig:Hamaker-r2} we report the same comparison with the NP diameter fixed to 4.0 nm. 
In a first
instance, we note that the minimum deep is less enhanced than the previous case: such a result
is apparently counterintuitive, since the increase of the particle size (and of the number of
beads) is expected to promote the attraction. But in this case the increase of the particle volume
is not fully compensated by the increase of the beads, this circumstance making the silica
nanoparticles less dense than before. Therefore, each bead experiments a weaker interaction and
the attractive well is less enhanced. In addition, it is worth noting that a repulsive contribution to
the total interaction comes from the coulombic potential: due to the presence of hydrogen atoms on the
most external shells, two NPs coming in close contact experience a significant electrostatic
repulsion. This effect is partially offset by the hydrogen bonds between oxygen and hydrogen atoms 
belonging to different NPs. However, for interparticle distances higher than 0.3 nm (see Ref.~\cite{Denicola:15}), 
the hydrogen bond can not take place, this giving rise to the shoulder observed in the PMF.
These effects are not observed for smaller silica NPs, since they are given by surface interactions and 
therefore are unfavoured if the number of atoms lying on the particles surface decreases.
By comparing simulation data with the Hamaker theory, we observe that
also in this case all essential features of the theoretical predictions 
(and, in particular, the
depth and the position of the attractive well) are nicely catched by simulated PMF,
except for the presence of the above said repulsive shoulder.
Such an agreement constitutes a further validation of the numerical procedure implemented
for calculating the PMF; at the same time, this finding is 
also indicative of a good trasferibility of the theory, since the latter appears to
accurately work regardless of the specific value of the NP size.
%
\subsection{Gold nanoparticles}
\begin{figure}
\begin{center}
\includegraphics[width=8.0cm,angle=0]{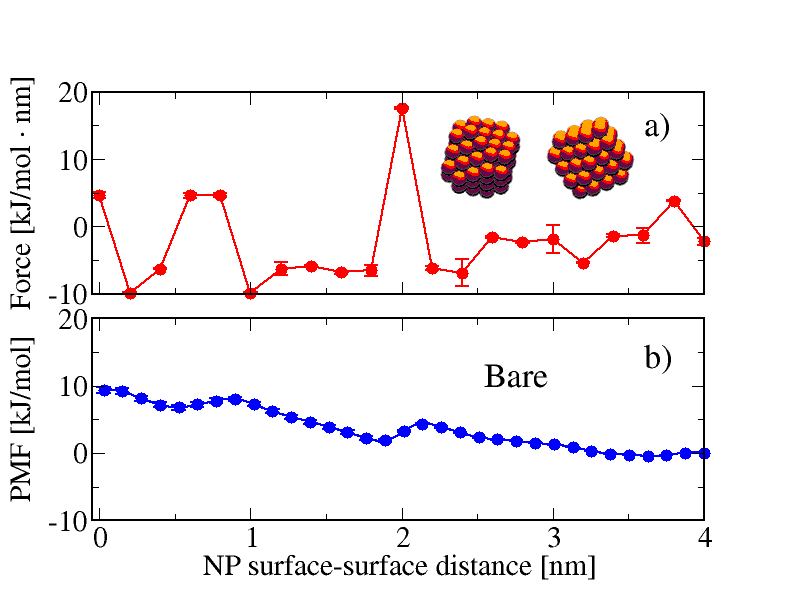} \\
\caption{Force (a) and PMF (b) between a pair of bare gold 
nanoparticles as a function of their mutual distance.}
\label{fig:gold-nak}
\end{center}
\end{figure}

Once assessed simulation results for silica NPs against Hamaker predictions, we 
now investigate the behavior of PMF between gold nanoparticles. In this case the
comparison with the theory can not be performed due to both the
non-spherical shape of the NPs and the presence of coated chains.
We first report the behaviors of force and PMF for a pair of bare gold NPs 
in panels (a) and (b) of Fig.~\ref{fig:gold-nak}, along with a pictorial 
view of the two NPs. The force appears quite noisy and lies in the range [-10 kJ/mol 
$\cdot$ nm; +20 kJ/mol $\cdot$ nm]; hence, it appears that the NPs are rather low 
interacting, as can be expected given the low number of atoms constituting a single
nanoparticle. Moreover, since a large part of such atoms lie on the surface of the NP
rather than in the core, the surface effects play a significant role, causing the
irregular behavior of the resulting force. By looking at the potential of mean force 
(panel b) we note that is shows a smoother trend and that it is repulsive in all the 
interparticle distance range, even if two very shallow minima are observed for NP surfaces placed at
0.53 and 1.90 nm. The overall behavior of the PMF is rather flat, this confirming that
the net interaction between the two NPs is low. Hence,   
in comparison with silica NPs, the emerging picture is quite different, since the
strong attraction previously observed for low interparticle distances has now disappeared. 
\begin{figure}
\begin{center}
\includegraphics[width=8.0cm,angle=0]{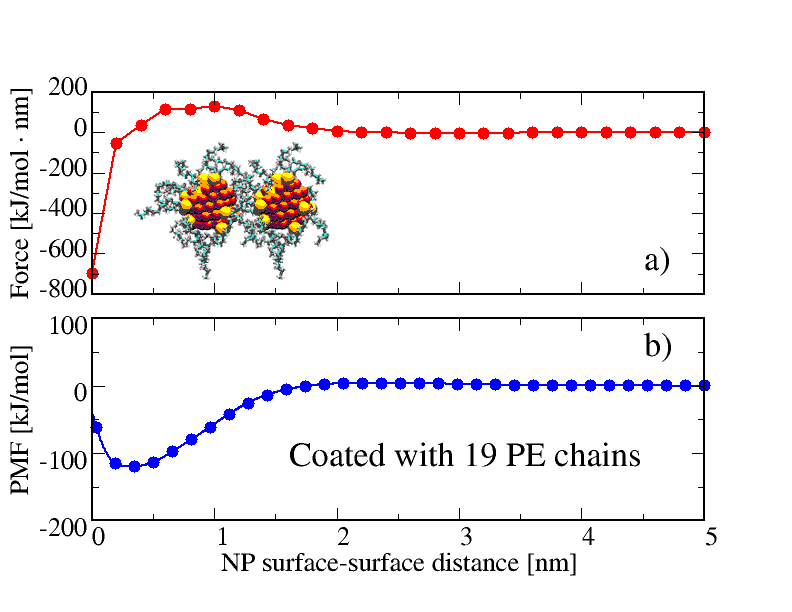} \\
\caption{Force (a) and PMF (b) between a pair of gold 
nanoparticles coated with 19 PE chains as a function of their mutual distance.
}
\label{fig:gold-19}
\end{center}
\end{figure}

Upon coating the gold nanoparticles with PE chains the behavior of the PMF is expected to be
modified, as a consequence of the interactions between chains. In Fig.~\ref{fig:gold-19} we report
force (panel a) and PMF (panel b) between a pair of gold NPs coated with 19 PE chains,
corresponding to a grafting density of $\rho_g=2.36$: in
comparison with the bare case, the force shows now a more regular and smooth behavior, the main
feature being the quick fall towards negative values in the range of short interparticle
distances. The potential of mean force exhibits a minimum placed at a distance of $\simeq$ 0.4 nm
and, interestingly, is still negative even if the two nanoparticles surfaces are in close contact.
This is likely due to the possibility for the chains to interpenetrate, since the relatively low
number (19) of chains coated on a particle leaves enough available space for the chains
belonging to the other particle. Such an effect is strongly dependent on the number of chains, 
{\it i.e.} on the grafting density $\rho_g$ and also on the distance between the nanoparticles.
A proper combination of these two parameters gives rise to the minimum observed in the PMF.

\begin{figure}
\begin{center}
\includegraphics[width=8.0cm,angle=0]{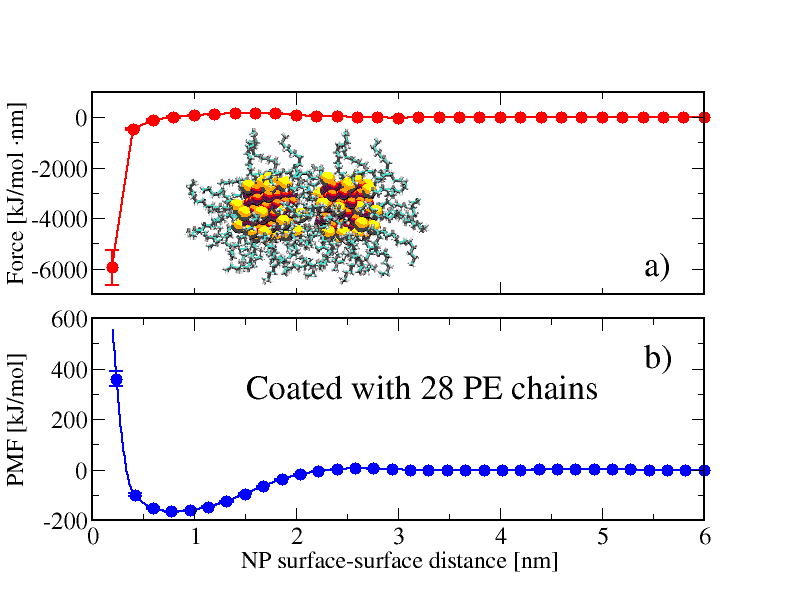} \\
\caption{Force (a) and PMF (b) between a pair of gold 
nanoparticles coated with 28 PE chains as a function of their mutual distance.}
\label{fig:gold-28}
\end{center}
\end{figure}

Results for gold nanoparticles coated with 28 PE chains, corresponding to a grafting
density of 3.48, are reported in Fig.~\ref{fig:gold-28}. The force (panel a)
is now strongly
negative for close-contact configurations, then showing a small and broad shoulder before
getting the asymptotic value for interparticle separations of $\simeq$ 2 nm. 
As a consequence, the PMF (panel b) is now significantly repulsive for 
close-contact configurations and
shows a well-defined attractive part with a minimum for interparticle distances of $\simeq$
0.8 nm. The overall behavior of the PMF could resemble to a Lennard-Jones potential, but 
in the present case the attractive well is quite broad, extending for $\simeq$ 1 nm.
The emerging picture suggests that with the increase of the grafting density there is also
an increase of the repulsion for very short ranges: this is due to the higher number of
chains that can not overlap, hence giving rise to repulsive contributions. When the two
gold nanoparticles are sligthly more distant, chains have enough space to interpenetrate,
thus generating an attractive interaction. Since there is a rather wide range of distances
where this interpenetration is possible, the minimum in the PMF appears broad and well-defined.

\begin{figure}
\begin{center}
\includegraphics[width=8.0cm,angle=0]{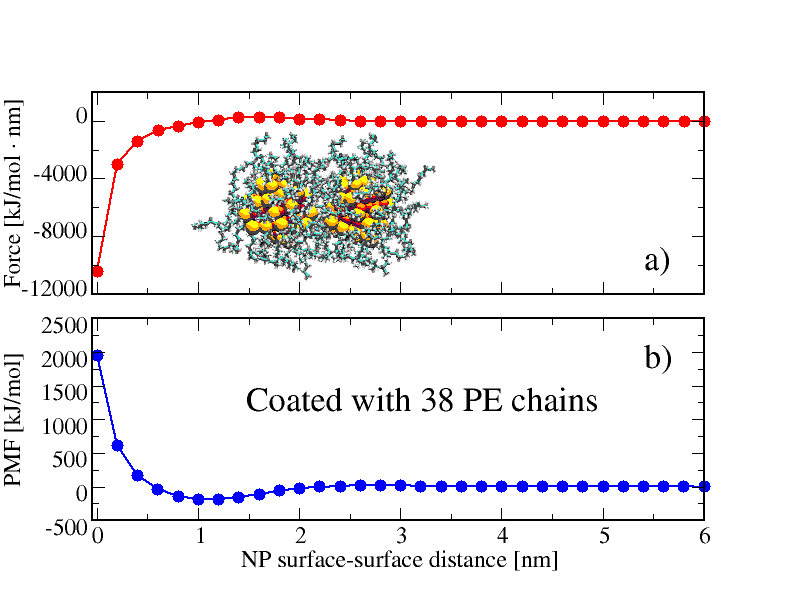} 
\caption{Force (a) and PMF (b) between a pair of gold 
nanoparticles coated with 38 PE chains as a function of their mutual distance.}
\label{fig:gold-38}
\end{center}
\end{figure}
\begin{figure}
\begin{center}
\includegraphics[width=8.0cm,angle=0]{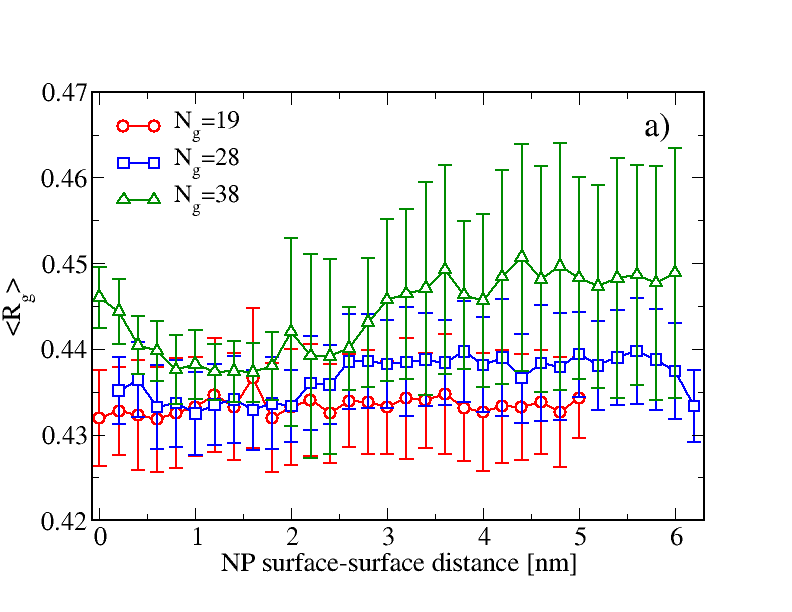} \\
\includegraphics[width=8.0cm,angle=0]{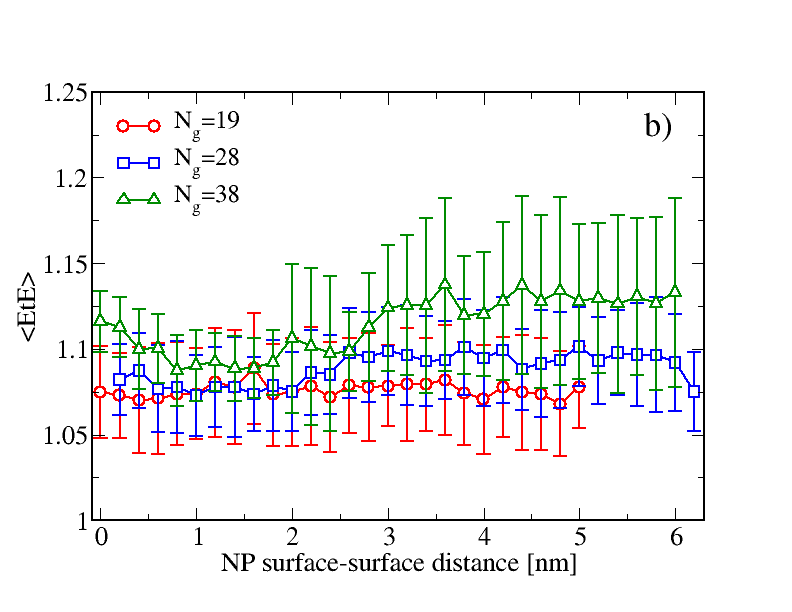} 
\caption{Average gyration radius (a) and average end-to-end distance (b) for PE chains with 
different $N_g$ as a function of the distance between NPs surfaces. Values of $N_g$ are in the
legends.}
\label{fig:rg}
\end{center}
\end{figure}
The case corresponding to the higher grafting density investigated in the present study 
($\rho_g=4.72$) is reported in Fig.~\ref{fig:gold-38}. A comparison with the previous case
shows that the force (panel a) is now remarkably more negative for very low interparticle separations;
on the other hand, the force  
attains its asymptotic value for surface distances of $\simeq 2$ nm, as observed for $\rho_g=3.48$. 
Overall, the behavior of the force is
similar to that previously observed, but for the strength of the force when the two NPs come in close
contact. As a consequence, the PMF (panel b) is now much repulsive for low interparticle separations and
shows a minimum of $\simeq$ -200 kJ/mol placed at a distance of $\simeq$ 1.2 nm. Upon comparing this 
behavior with the cases of lower grafting densities, we note that the close-contact value of the PMF
remarkably increases with $\rho_g$; the position of the minimum is also affected by the grafting density, going
from 0.4 nm for $\rho_g=2.36$ till to 1.2 nm for $\rho_g=4.72$. The depth of the minimum is indeed 
unchanged when going from $\rho_g=3.48$ to $\rho_g=4.72$. All these features can be explained in terms
of chain interpenetration: upon increasing $\rho_g$, chains belonging to different NPs are progressively
repelled from each other, this giving rise to the increasing repulsion observed in the PMF; on the other
hand, there is a preferred NP-NP distance where the chains can be more easily arranged, hence giving rise
to a minimum in the PMF, whose position is in turn dependent on $\rho_g$.

\begin{figure}
\begin{center}
\includegraphics[width=8.0cm,angle=0]{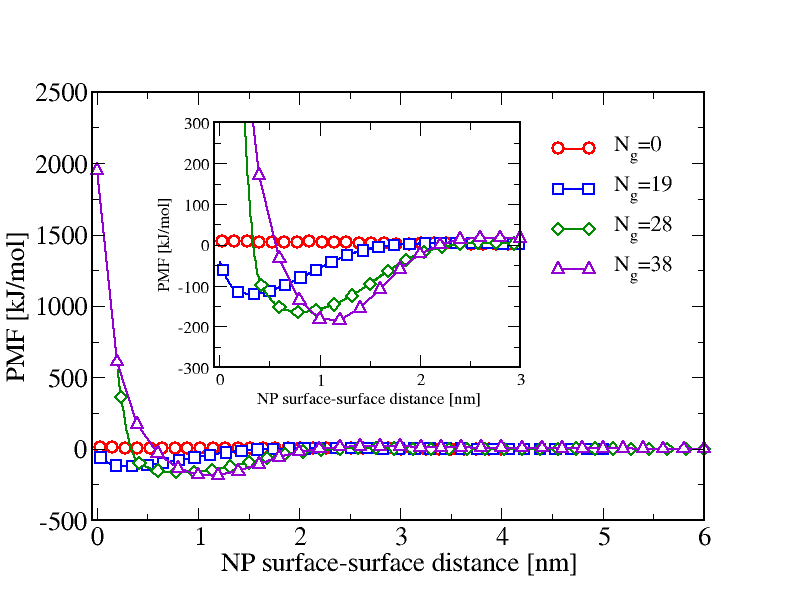} 
\caption{Comparison between PMF of a pair of gold 
nanoparticles for different values of $N_g$ (in the legend) as a function of their 
mutual distance. The behavior for low interparticle distances is highlighted in the inset.}
\label{fig:gold-all}
\end{center}
\end{figure}
More insight into the extension of PE chains and their dependence on the grafting density 
can be gained by investigating their local structure.
In Fig.~\ref{fig:rg} we report the average gyration
radius $\langle R_g \rangle$ (panel a) and the average end-to-end distance $\langle EtE \rangle$
(panel b) for all the investigated values of $N_g$. 
As a first instance, we note that 
both $\langle R_g \rangle$ and $\langle EtE \rangle$ shows a clear dependence on $N_g$: as visible,
upon increasing the number of coating chains, values of 
$\langle R_g \rangle$ and $\langle EtE \rangle$ increase in turn, this being particularly
apparent for $N_g=38$. This finding is a clear indication that for high $N_g$ (and hence for
high grafting densities) the chains are more stretched, since the available space is reduced and 
they are forced to assume elongated configurations.
On the other hand, $\langle R_g \rangle$ and $\langle EtE \rangle$ shows also a slight dependence
on the interparticle separation, even if such a dependence is less enhanced than that on $N_g$: in
fact, for $N_g=28$ and, in particular, for $N_g=38$, we note the presence of a shift for 
a NP-NP distance of $\simeq 2.5$ nm. In correspondence of this distance, both $\langle R_g \rangle$ and 
$\langle EtE \rangle$ jump to higher values, thus signalling a point where the coated chains belonging to
a NP stop to be compressed by the chains belonging to the other NP, hence returning to their
unperturbed size. By comparing this behavior with the PMF reported in Fig.~\ref{fig:gold-28} and
Fig.~\ref{fig:gold-38}, one can note that the shift appears for the same interparticle separation
where the PMF vanishes, in agreement with the recovering of the unperturbed state of the chains.

A summarizing comparison between all the PMF calculated for gold NPs is given in Fig.~\ref{fig:gold-all}: 
in particular, the increase of the repulsion strength for low interparticle distance is clearly visible.
In the inset, the region around PMF minima is magnified: here, the progressive shift of the minimum
towards higher values of the interparticle separation is enhanced, in agreement with the increase 
of $\langle R_g \rangle$ and $\langle EtE \rangle$ previously discussed.

\section{Conclusions}
In the present work we have investigated the behavior of the potential of mean
force (PMF) between atomistic silica or gold nanoparticles (NPs). By performing
GROMACS	molecular simulations in the canonical ensemble, we have calculated the net forces between a 
pair of such particles, hence obtaining the PMF through an integration over
their mutual distances. 
In the case of silica nanoparticles, the effects
due to the particle size have been taken into account upon calculating the
PMF for particles of 2.5 and 4 nm of diameter. No significant discrepancies 
between the two cases have been observed, the only remarkable difference being the
appearance of a peak in the PMF between the two larger particles. Such an 
effect is likely due to surface interactions given by the formation and 
breaking of hydrogen bonds. In addition, we have performed a comparison between
simulation data and an analytical theory
due to Hamaker and obtained by employing the Lennard-Jones parameters of 
oxygen and silicon atoms. As a result, a good agreement between theory and simulations
has been found, this suggesting 
that for bare spherical particles made by a large number of atoms the overall behavior
of the PMF can be well approximated by means of a combination of site-site
Lennard-Jones potentials. 
We have then investigated the behavior of PMF between
gold nanoparticles, which have been
considered in a first instance bare and then coated with an increasing number
of polyethilene chains; in such a way it is possible to detect the effect of the grafting density
$\rho_g$ on the PMF. We have found that bare gold NPs experience little
interactions and surface effects are dominant; upon coating the particles
with polyethilene chains, the profile of the PMF is deeply modified, with the
appearance of a short-range attraction characterized by a large attractive
well. Overall, the resulting PMF appears quite similar to a Lennard-Jones
potential, but for the attractive well that appears broader, extending for
distances of $\simeq$ 1 nm. An high repulsive interaction is detected for low
interparticle distances and high grafting densities.

The present work can be considered a preliminary and necessary study suited
to fully characterize effective interactions in high-density coarse-grained polymer nanocomposites, 
where the knowledge of atomistic potentials constitute a crucial issue that need to be
properly taken into account. Such studies will constitute the main target of a forthcoming
investigation. 

\section{Acknowledgements}
The computing resources and the related technical support used for this work 
have been provided by CRESCO/ENEAGRID High Performance Computing infrastructure
and its staff~\cite{Cresco}. CRESCO/ENEAGRID High Performance Computing 
infrastructure is funded by ENEA, the Italian National Agency for New 
Technologies, Energy and Sustainable Economic Development and by Italian and 
European research programmes, see http://www.cresco.enea.it/english for 
information. G.~Muna\`o and G.~Milano acknowledge financial support from 
the project PRIN-MIUR 2015-2016.

\section{Author contribution statement}
Giuseppe Milano suggested the research topic and coordinated the preparation
of the manuscript. Gianmarco Muna\`o wrote the paper and performed numerical
simulations along with Andrea Correa. Gianmarco Muna\`o and Antonio Pizzirusso 
performed the structural analysis. All authors contributed to the 
critical reading of the manuscript.


\begin{mcitethebibliography}{43}
\providecommand*{\natexlab}[1]{#1}
\providecommand*{\mciteSetBstSublistMode}[1]{}
\providecommand*{\mciteSetBstMaxWidthForm}[2]{}
\providecommand*{\mciteBstWouldAddEndPuncttrue}
  {\def\EndOfBibitem{\unskip.}}
\providecommand*{\mciteBstWouldAddEndPunctfalse}
  {\let\EndOfBibitem\relax}
\providecommand*{\mciteSetBstMidEndSepPunct}[3]{}
\providecommand*{\mciteSetBstSublistLabelBeginEnd}[3]{}
\providecommand*{\EndOfBibitem}{}
\mciteSetBstSublistMode{f}
\mciteSetBstMaxWidthForm{subitem}
{(\emph{\alph{mcitesubitemcount}})}
\mciteSetBstSublistLabelBeginEnd{\mcitemaxwidthsubitemform\space}
{\relax}{\relax}

\bibitem[Akcora \emph{et~al.}(2009)Akcora, Liu, Kumar, Moll, Li, Benicewicz,
  Schadler, Acechin, Panagiotopoulos, Pyramitsyn, Ganesan, Ilavsky,
  Thiyagarajan, Colby, and Douglas]{Akcora:09}
P.~Akcora, H.~Liu, S.~K. Kumar, J.~Moll, Y.~Li, B.~C. Benicewicz, L.~S.
  Schadler, D.~Acechin, A.~Z. Panagiotopoulos, V.~Pyramitsyn, V.~Ganesan,
  J.~Ilavsky, P.~Thiyagarajan, R.~H. Colby and J.~F. Douglas, \emph{Nat.
  Mater.}, 2009, \textbf{8}, 354\relax
\mciteBstWouldAddEndPuncttrue
\mciteSetBstMidEndSepPunct{\mcitedefaultmidpunct}
{\mcitedefaultendpunct}{\mcitedefaultseppunct}\relax
\EndOfBibitem
\bibitem[Kawada \emph{et~al.}(2017)Kawada, Fujimoto, Yoshii, and
  Okazaki]{Kawada:17}
S.~Kawada, K.~Fujimoto, N.~Yoshii and S.~Okazaki, \emph{J. Chem. Phys.}, 2017,
  \textbf{147}, 084903\relax
\mciteBstWouldAddEndPuncttrue
\mciteSetBstMidEndSepPunct{\mcitedefaultmidpunct}
{\mcitedefaultendpunct}{\mcitedefaultseppunct}\relax
\EndOfBibitem
\bibitem[Baran and Sokolowski(2017)]{Baran:17}
L.~Baran and S.~Sokolowski, \emph{J. Chem. Phys.}, 2017, \textbf{147},
  044903\relax
\mciteBstWouldAddEndPuncttrue
\mciteSetBstMidEndSepPunct{\mcitedefaultmidpunct}
{\mcitedefaultendpunct}{\mcitedefaultseppunct}\relax
\EndOfBibitem
\bibitem[Akcora \emph{et~al.}(2010)Akcora, Kumar, Moll, Lewis, Schadler, Li,
  Benicewicz, Sandy, Narayanan, Ilavsky, Thiyagarajan, Colby, and
  Douglas]{Akcora:10}
P.~Akcora, S.~K. Kumar, J.~Moll, S.~Lewis, L.~S. Schadler, Y.~Li, B.~C.
  Benicewicz, A.~Sandy, S.~Narayanan, J.~Ilavsky, P.~Thiyagarajan, R.~H. Colby
  and J.~F. Douglas, \emph{Macromolecules}, 2010, \textbf{43}, 1003\relax
\mciteBstWouldAddEndPuncttrue
\mciteSetBstMidEndSepPunct{\mcitedefaultmidpunct}
{\mcitedefaultendpunct}{\mcitedefaultseppunct}\relax
\EndOfBibitem
\bibitem[Han and Fina(2011)]{Han:11}
Z.~Han and A.~Fina, \emph{Prog. Polym. Sci.}, 2011, \textbf{36}, 914\relax
\mciteBstWouldAddEndPuncttrue
\mciteSetBstMidEndSepPunct{\mcitedefaultmidpunct}
{\mcitedefaultendpunct}{\mcitedefaultseppunct}\relax
\EndOfBibitem
\bibitem[Kim \emph{et~al.}(2012)Kim, Yang, and Green]{Kim:12}
J.~Kim, H.~Yang and P.~F. Green, \emph{Langmuir}, 2012, \textbf{28}, 9735\relax
\mciteBstWouldAddEndPuncttrue
\mciteSetBstMidEndSepPunct{\mcitedefaultmidpunct}
{\mcitedefaultendpunct}{\mcitedefaultseppunct}\relax
\EndOfBibitem
\bibitem[Borukhov and Leibler(2002)]{Leibler:02}
I.~Borukhov and L.~Leibler, \emph{Macromolecules}, 2002, \textbf{35},
  5171\relax
\mciteBstWouldAddEndPuncttrue
\mciteSetBstMidEndSepPunct{\mcitedefaultmidpunct}
{\mcitedefaultendpunct}{\mcitedefaultseppunct}\relax
\EndOfBibitem
\bibitem[Kumar \emph{et~al.}(2013)Kumar, Jouault, Benicewicz, and
  Neely]{Kumar:13}
S.~K. Kumar, N.~Jouault, B.~Benicewicz and T.~Neely, \emph{Macromolecules},
  2013, \textbf{46}, 3199\relax
\mciteBstWouldAddEndPuncttrue
\mciteSetBstMidEndSepPunct{\mcitedefaultmidpunct}
{\mcitedefaultendpunct}{\mcitedefaultseppunct}\relax
\EndOfBibitem
\bibitem[Chevigny \emph{et~al.}(2011)Chevigny, Dalmas, {Di Cola}, Gigmes,
  Bertin, Boue, and Jestin]{Chevigny:11}
C.~Chevigny, F.~Dalmas, E.~{Di Cola}, D.~Gigmes, D.~Bertin, F.~Boue and
  J.~Jestin, \emph{Macromolecules}, 2011, \textbf{44}, 122\relax
\mciteBstWouldAddEndPuncttrue
\mciteSetBstMidEndSepPunct{\mcitedefaultmidpunct}
{\mcitedefaultendpunct}{\mcitedefaultseppunct}\relax
\EndOfBibitem
\bibitem[You \emph{et~al.}(2017)You, Yu, and Zhou]{You:17}
W.~You, W.~Yu and C.~Zhou, \emph{Soft Matter}, 2017, \textbf{13}, 4088\relax
\mciteBstWouldAddEndPuncttrue
\mciteSetBstMidEndSepPunct{\mcitedefaultmidpunct}
{\mcitedefaultendpunct}{\mcitedefaultseppunct}\relax
\EndOfBibitem
\bibitem[Martin \emph{et~al.}(2013)Martin, Dodd, and Jayaraman]{Martin:13}
T.~B. Martin, P.~M. Dodd and A.~Jayaraman, \emph{Phys. Rev. Lett.}, 2013,
  \textbf{110}, 018301\relax
\mciteBstWouldAddEndPuncttrue
\mciteSetBstMidEndSepPunct{\mcitedefaultmidpunct}
{\mcitedefaultendpunct}{\mcitedefaultseppunct}\relax
\EndOfBibitem
\bibitem[Ganesan and Jayaraman(2014)]{Ganesan:14}
V.~Ganesan and A.~Jayaraman, \emph{Soft Matter}, 2014, \textbf{10}, 13\relax
\mciteBstWouldAddEndPuncttrue
\mciteSetBstMidEndSepPunct{\mcitedefaultmidpunct}
{\mcitedefaultendpunct}{\mcitedefaultseppunct}\relax
\EndOfBibitem
\bibitem[Meng \emph{et~al.}(2012)Meng, Kumar, Lane, and Grest]{Meng:12}
D.~Meng, S.~K. Kumar, J.~M.~D. Lane and G.~S. Grest, \emph{Soft Matter}, 2012,
  \textbf{8}, 5002\relax
\mciteBstWouldAddEndPuncttrue
\mciteSetBstMidEndSepPunct{\mcitedefaultmidpunct}
{\mcitedefaultendpunct}{\mcitedefaultseppunct}\relax
\EndOfBibitem
\bibitem[Karatrantos \emph{et~al.}(2017)Karatrantos, Composto, Winey, and
  Clarke]{Karatrantos:17}
A.~Karatrantos, R.~J. Composto, K.~I. Winey and N.~Clarke, \emph{J. Chem.
  Phys.}, 2017, \textbf{146}, 203331\relax
\mciteBstWouldAddEndPuncttrue
\mciteSetBstMidEndSepPunct{\mcitedefaultmidpunct}
{\mcitedefaultendpunct}{\mcitedefaultseppunct}\relax
\EndOfBibitem
\bibitem[Schweizer and Curro(1987)]{Curro1}
K.~S. Schweizer and J.~G. Curro, \emph{Phys. Rev. Lett.}, 1987, \textbf{58},
  246\relax
\mciteBstWouldAddEndPuncttrue
\mciteSetBstMidEndSepPunct{\mcitedefaultmidpunct}
{\mcitedefaultendpunct}{\mcitedefaultseppunct}\relax
\EndOfBibitem
\bibitem[Curro and Schweizer(1987)]{Curro2}
J.~G. Curro and K.~S. Schweizer, \emph{Macromolecules}, 1987, \textbf{20},
  1928\relax
\mciteBstWouldAddEndPuncttrue
\mciteSetBstMidEndSepPunct{\mcitedefaultmidpunct}
{\mcitedefaultendpunct}{\mcitedefaultseppunct}\relax
\EndOfBibitem
\bibitem[Hooper \emph{et~al.}(2004)Hooper, Schweizer, Desai, Koshy, and
  Keblinski]{Hooper:04}
J.~B. Hooper, K.~S. Schweizer, T.~G. Desai, R.~Koshy and P.~Keblinski, \emph{J.
  Chem. Phys.}, 2004, \textbf{121}, 6986\relax
\mciteBstWouldAddEndPuncttrue
\mciteSetBstMidEndSepPunct{\mcitedefaultmidpunct}
{\mcitedefaultendpunct}{\mcitedefaultseppunct}\relax
\EndOfBibitem
\bibitem[Jayaraman and Schweizer(2009)]{Jayaraman:09}
A.~Jayaraman and K.~S. Schweizer, \emph{Macromolecules}, 2009, \textbf{42},
  8423\relax
\mciteBstWouldAddEndPuncttrue
\mciteSetBstMidEndSepPunct{\mcitedefaultmidpunct}
{\mcitedefaultendpunct}{\mcitedefaultseppunct}\relax
\EndOfBibitem
\bibitem[Allegra \emph{et~al.}(2008)Allegra, Raos, and Vacatello]{Raos:08}
G.~Allegra, G.~Raos and M.~Vacatello, \emph{Prog. Polym. Sci.}, 2008,
  \textbf{33}, 683\relax
\mciteBstWouldAddEndPuncttrue
\mciteSetBstMidEndSepPunct{\mcitedefaultmidpunct}
{\mcitedefaultendpunct}{\mcitedefaultseppunct}\relax
\EndOfBibitem
\bibitem[Ganesan and Jayaraman(2014)]{Ganesan:10}
V.~Ganesan and A.~Jayaraman, \emph{Soft Matter}, 2014, \textbf{10}, 13\relax
\mciteBstWouldAddEndPuncttrue
\mciteSetBstMidEndSepPunct{\mcitedefaultmidpunct}
{\mcitedefaultendpunct}{\mcitedefaultseppunct}\relax
\EndOfBibitem
\bibitem[Karatrantos \emph{et~al.}(2016)Karatrantos, Clarke, and
  Kr{\"o}ger]{Karatrantos:16}
A.~Karatrantos, N.~Clarke and M.~Kr{\"o}ger, \emph{Polymer Reviews}, 2016,
  \textbf{56}, 385\relax
\mciteBstWouldAddEndPuncttrue
\mciteSetBstMidEndSepPunct{\mcitedefaultmidpunct}
{\mcitedefaultendpunct}{\mcitedefaultseppunct}\relax
\EndOfBibitem
\bibitem[Kumar \emph{et~al.}(2017)Kumar, Ganesan, and Riggleman]{Kumar:17}
S.~K. Kumar, V.~Ganesan and R.~A. Riggleman, \emph{J. Chem. Phys.}, 2017,
  \textbf{147}, 020901\relax
\mciteBstWouldAddEndPuncttrue
\mciteSetBstMidEndSepPunct{\mcitedefaultmidpunct}
{\mcitedefaultendpunct}{\mcitedefaultseppunct}\relax
\EndOfBibitem
\bibitem[Cerd{\`a} \emph{et~al.}(2003)Cerd{\`a}, Sintes, and Toral]{Cerda:03}
J.~J. Cerd{\`a}, T.~Sintes and R.~Toral, \emph{Macromolecules}, 2003,
  \textbf{36}, 1407\relax
\mciteBstWouldAddEndPuncttrue
\mciteSetBstMidEndSepPunct{\mcitedefaultmidpunct}
{\mcitedefaultendpunct}{\mcitedefaultseppunct}\relax
\EndOfBibitem
\bibitem[Smith \emph{et~al.}(2003)Smith, Bedrov, and Smith]{Smith:03}
J.~S. Smith, D.~Bedrov and G.~D. Smith, \emph{Compos. Sci. Technol.}, 2003,
  \textbf{63}, 1599\relax
\mciteBstWouldAddEndPuncttrue
\mciteSetBstMidEndSepPunct{\mcitedefaultmidpunct}
{\mcitedefaultendpunct}{\mcitedefaultseppunct}\relax
\EndOfBibitem
\bibitem[Marla and Meredith(2006)]{Marla:06}
K.~T. Marla and J.~C. Meredith, \emph{J. Chem. Theory Comput.}, 2006,
  \textbf{2}, 1624\relax
\mciteBstWouldAddEndPuncttrue
\mciteSetBstMidEndSepPunct{\mcitedefaultmidpunct}
{\mcitedefaultendpunct}{\mcitedefaultseppunct}\relax
\EndOfBibitem
\bibitem[Smith and Bedrov(2009)]{Smith:09}
G.~D. Smith and D.~Bedrov, \emph{Langmuir}, 2009, \textbf{25}, 11239\relax
\mciteBstWouldAddEndPuncttrue
\mciteSetBstMidEndSepPunct{\mcitedefaultmidpunct}
{\mcitedefaultendpunct}{\mcitedefaultseppunct}\relax
\EndOfBibitem
\bibitem[{Lo Verso} \emph{et~al.}(2011){Lo Verso}, Yelash, Egorov, and
  Binder]{Loverso:11}
F.~{Lo Verso}, L.~Yelash, S.~A. Egorov and K.~Binder, \emph{J. Chem. Phys.},
  2011, \textbf{135}, 214902\relax
\mciteBstWouldAddEndPuncttrue
\mciteSetBstMidEndSepPunct{\mcitedefaultmidpunct}
{\mcitedefaultendpunct}{\mcitedefaultseppunct}\relax
\EndOfBibitem
\bibitem[Milano and Kawakatsu(2009)]{Milano:09}
G.~Milano and T.~Kawakatsu, \emph{J. Chem. Phys.}, 2009, \textbf{130},
  214106\relax
\mciteBstWouldAddEndPuncttrue
\mciteSetBstMidEndSepPunct{\mcitedefaultmidpunct}
{\mcitedefaultendpunct}{\mcitedefaultseppunct}\relax
\EndOfBibitem
\bibitem[Milano and Kawakatsu(2010)]{Milano:10}
G.~Milano and T.~Kawakatsu, \emph{J. Chem. Phys.}, 2010, \textbf{133},
  214102\relax
\mciteBstWouldAddEndPuncttrue
\mciteSetBstMidEndSepPunct{\mcitedefaultmidpunct}
{\mcitedefaultendpunct}{\mcitedefaultseppunct}\relax
\EndOfBibitem
\bibitem[Hess \emph{et~al.}(2008)Hess, Kutzner, {Van del Spoel}, and
  Lindahl]{Gromacs:08}
B.~Hess, C.~Kutzner, D.~{Van del Spoel} and E.~Lindahl, \emph{J. Chem. Theor.
  Comput.}, 2008, \textbf{4}, 435\relax
\mciteBstWouldAddEndPuncttrue
\mciteSetBstMidEndSepPunct{\mcitedefaultmidpunct}
{\mcitedefaultendpunct}{\mcitedefaultseppunct}\relax
\EndOfBibitem
\bibitem[Hamaker(1937)]{Hamaker:37}
H.~C. Hamaker, \emph{Physica}, 1937, \textbf{4}, 1058\relax
\mciteBstWouldAddEndPuncttrue
\mciteSetBstMidEndSepPunct{\mcitedefaultmidpunct}
{\mcitedefaultendpunct}{\mcitedefaultseppunct}\relax
\EndOfBibitem
\bibitem[Ndoro \emph{et~al.}(2011)Ndoro, Voyiatzis, Ghanbari, Theodorou,
  B{\"o}hm, and M{\"u}ller-Plathe]{Ndoro:11}
T.~V.~M. Ndoro, E.~Voyiatzis, A.~Ghanbari, D.~N. Theodorou, M.~C. B{\"o}hm and
  F.~M{\"u}ller-Plathe, \emph{Macromolecules}, 2011, \textbf{44}, 2316\relax
\mciteBstWouldAddEndPuncttrue
\mciteSetBstMidEndSepPunct{\mcitedefaultmidpunct}
{\mcitedefaultendpunct}{\mcitedefaultseppunct}\relax
\EndOfBibitem
\bibitem[Eslami \emph{et~al.}(2013)Eslami, Rahimi, and
  M{\"u}ller-Plathe]{Eslami:13}
H.~Eslami, M.~Rahimi and F.~M{\"u}ller-Plathe, \emph{Macromolecules}, 2013,
  \textbf{46}, 8680\relax
\mciteBstWouldAddEndPuncttrue
\mciteSetBstMidEndSepPunct{\mcitedefaultmidpunct}
{\mcitedefaultendpunct}{\mcitedefaultseppunct}\relax
\EndOfBibitem
\bibitem[Rai \emph{et~al.}(2004)Rai, Sathish, Malhotra, Pradip, and
  Ayappa]{Rai:04}
B.~Rai, P.~Sathish, C.~P. Malhotra, Pradip and K.~G. Ayappa, \emph{Langmuir},
  2004, \textbf{20}, 3138\relax
\mciteBstWouldAddEndPuncttrue
\mciteSetBstMidEndSepPunct{\mcitedefaultmidpunct}
{\mcitedefaultendpunct}{\mcitedefaultseppunct}\relax
\EndOfBibitem
\bibitem[Milano \emph{et~al.}(2011)Milano, Santangelo, Ragone, Cavallo, and {Di
  Matteo}]{Milano-Gold}
G.~Milano, G.~Santangelo, F.~Ragone, L.~Cavallo and A.~{Di Matteo}, \emph{J.
  Phys. Chem. C}, 2011, \textbf{115}, 15154\relax
\mciteBstWouldAddEndPuncttrue
\mciteSetBstMidEndSepPunct{\mcitedefaultmidpunct}
{\mcitedefaultendpunct}{\mcitedefaultseppunct}\relax
\EndOfBibitem
\bibitem[Tironi \emph{et~al.}(1995)Tironi, Sperb, Smith, and {van
  Gunsteren}]{Tironi:95}
I.~G. Tironi, R.~Sperb, P.~E. Smith and W.~F. {van Gunsteren}, \emph{J. Chem.
  Phys.}, 1995, \textbf{102}, 5451\relax
\mciteBstWouldAddEndPuncttrue
\mciteSetBstMidEndSepPunct{\mcitedefaultmidpunct}
{\mcitedefaultendpunct}{\mcitedefaultseppunct}\relax
\EndOfBibitem
\bibitem[Kirkwood(1935)]{Kirkwood:35}
J.~G. Kirkwood, \emph{J. Chem. Phys.}, 1935, \textbf{3}, 300\relax
\mciteBstWouldAddEndPuncttrue
\mciteSetBstMidEndSepPunct{\mcitedefaultmidpunct}
{\mcitedefaultendpunct}{\mcitedefaultseppunct}\relax
\EndOfBibitem
\bibitem[{De Nicola} \emph{et~al.}(2017){De Nicola}, Correa, Milano, {La
  Manna}, Musto, Mensitieri, and Scherillo]{DeNicola-JPCB}
A.~{De Nicola}, A.~Correa, G.~Milano, P.~{La Manna}, P.~Musto, G.~Mensitieri
  and G.~Scherillo, \emph{J. Phys. Chem. B}, 2017, \textbf{121}, 3162\relax
\mciteBstWouldAddEndPuncttrue
\mciteSetBstMidEndSepPunct{\mcitedefaultmidpunct}
{\mcitedefaultendpunct}{\mcitedefaultseppunct}\relax
\EndOfBibitem
\bibitem[Berendsen \emph{et~al.}(1984)Berendsen, Postma, Gunsteren, {Di Nola},
  and Haak]{Berendsen:84}
H.~J.~C. Berendsen, J.~P.~M. Postma, W.~F.~V. Gunsteren, A.~{Di Nola} and J.~R.
  Haak, \emph{J. Chem. Phys.}, 1984, \textbf{81}, 3684\relax
\mciteBstWouldAddEndPuncttrue
\mciteSetBstMidEndSepPunct{\mcitedefaultmidpunct}
{\mcitedefaultendpunct}{\mcitedefaultseppunct}\relax
\EndOfBibitem
\bibitem[Martinez \emph{et~al.}(2009)Martinez, Andrade, Birgin, and
  Martinez]{Packmol}
L.~Martinez, R.~Andrade, E.~G. Birgin and J.~M. Martinez, \emph{J. Comput.
  Chem.}, 2009, \textbf{30}, 2157\relax
\mciteBstWouldAddEndPuncttrue
\mciteSetBstMidEndSepPunct{\mcitedefaultmidpunct}
{\mcitedefaultendpunct}{\mcitedefaultseppunct}\relax
\EndOfBibitem
\bibitem[Everaers and Ejtehadi(2003)]{Everaers:03}
R.~Everaers and M.~R. Ejtehadi, \emph{Phys. Rev. E}, 2003, \textbf{67},
  041710\relax
\mciteBstWouldAddEndPuncttrue
\mciteSetBstMidEndSepPunct{\mcitedefaultmidpunct}
{\mcitedefaultendpunct}{\mcitedefaultseppunct}\relax
\EndOfBibitem
\bibitem[{De Nicola} \emph{et~al.}(2015){De Nicola}, Avolio, {Della Monica},
  Gentile, Cocca, Capacchione, Errico, and Milano]{Denicola:15}
A.~{De Nicola}, R.~Avolio, F.~{Della Monica}, G.~Gentile, M.~Cocca,
  C.~Capacchione, M.~E. Errico and G.~Milano, \emph{RSC Advances}, 2015,
  \textbf{5}, 71336\relax
\mciteBstWouldAddEndPuncttrue
\mciteSetBstMidEndSepPunct{\mcitedefaultmidpunct}
{\mcitedefaultendpunct}{\mcitedefaultseppunct}\relax
\EndOfBibitem
\bibitem[Ponti \emph{et~al.}(2014)Ponti\emph{et~al.}]{Cresco}
G.~Ponti \emph{et~al.}, \emph{Proceedings of the 2014 International Conference
  on High Performance Computing and Simulation}, 2014, \textbf{6903807},
  1030\relax
\mciteBstWouldAddEndPuncttrue
\mciteSetBstMidEndSepPunct{\mcitedefaultmidpunct}
{\mcitedefaultendpunct}{\mcitedefaultseppunct}\relax
\EndOfBibitem
\end{mcitethebibliography}

\providecommand*{\mcitethebibliography}{\thebibliography}
\csname @ifundefined\endcsname{endmcitethebibliography}
{\let\endmcitethebibliography\endthebibliography}{}

\end{document}